A general, simple, robust method to account for measurement error when analyzing data with an internal validation subsample

Walter K Kremers, PhD

October 20, 2021

Affiliation:   Department of Quantitative Sciences

Mayo Clinic

200 First St. SW

Rochester MN 55905, USA

Email:   kremers.walter@mayo.edu

Word count abstract: 249

Word count main body: ~ 5,000

This work was funded by the National Institutes of Health, National Institute of Arthritis and Musculoskeletal and Skin Diseases grant R01AR73147 and P30AR76312.  The author has received research funding unrelated to this study from Biogen, Roche and AstraZeneca.

Clinical data for an example were provided by Hilal Maradit Kremers.

**ABSTRACT**

**Background:** Measurement errors in terms of quantification or classification frequently occur in epidemiologic data and can strongly impact inference. Measurement errors may occur when ascertaining, recording or extracting data. Although the effects of measurement errors can be severe and are well described, simple straight forward general analytic solutions are not readily available for statistical analysis and measurement error is frequently not acknowledged or accounted for. Generally, to account for measurement error requires some data where we can observe the variables once with and once without error, to establish the relationship between the two.

**Methods:** Here we describe a general method accounting for measurement error in outcome and/or predictor variables for the parametric regression setting when there is a validation subsample where variables are measured once with and once without error. The method does not describe and thus does not depend on the particular relation between the variables measured with and without error, and is generally robust to the type of measurement error, for example nondifferential, differential or Berkson errors.

**Results:** Simulation studies show how the method reduces bias compared to models based upon variables measured with error alone and reduces variances compared to models based upon the variables measured without error in the validation subsample alone.

**Conclusion:** The proposed estimator has favorable properties in terms of bias and variance, is easily derived empirically, and is robust to different types of measurement error. This method should be a valuable tool in the analysis of data with measurement error.

**Key Messages**

- Measurement errors do not just attenuate effect estimates but may bias estimates upward or downward, can bias estimates for terms not affected by measurement error, and should not be ignored.
- Statistical approaches to mitigate the effect of measurement error are many but there are no generally applicable methods with practical user friendly software.
- We present a general method for the regression setting which can be used when there exists a validation subsample. This method is generally applicable and robust to different types of measurement error including non-differential, differential and Berkson errors.
- Open source user friendly software is made available making the method accessible to those who understand measurement error but do not have extensive programming experience.

**INTRODUCTION**

Measurement errors in terms of quantification or classification frequently occur in epidemiologic data and can strongly impact inference. Errors may occur when data are ascertained, that is measured or classified at the time of patient contact, when recorded, that is entered into a database, or when extracted, for example when read from a data source like a health record. Increasingly artificial intelligence is used to extract information form the electronic health record, for example using recursive neural networks to determine symptoms and diagnosis from clinical notes, or using convolution networks to identify pathologies and diagnoses based from radiographic, biopsy and photographic images as well as EKG, EEG and EMG tracings. These new technologies are extremely valuable allowing standardized assimilation of data for large numbers of patients, but may be subject to error when compared to a manual review by an experienced medical abstractor or physician, bringing new ways for errors to enter into data for analysis.

Although the effects of measurement errors can be severe and are well described, this issue is frequently not acknowledged or accounted for[1,2,3,4]. Often times it is misunderstood that the worst impact of measurement error is that parameter estimates for the terms measured with error may be somewhat attenuated, but in fact measurement error may also inflate parameter estimates and effect estimates for terms not affected by measurement error[1,3]. This and other misunderstandings are often given to justify ignoring measurement error at the time of statistical analyses[3,4].

Measurement errors in data can occur in a myriad of ways involving quantification or classification, in either outcome (predicted) or predictor (explanatory) variables, or both.  Measurement errors may be simpler in form like in the case of non-differential error.  Roughly, errors in predictors are non-differential if the model outcome is independent of the measurement errors, and only dependent on the reference or true variables, and errors in the outcome are non-differential if the errors are independent of the predictor variables[3].  If there are dependencies, then the errors are described as differential.  Differential errors can occur when a person is in a high risk group (predictor) for a disease (outcome) and there is greater scrutiny to correctly capture, record and clearly document the outcome variable. Alternatively, when a person is found to have a bad outcome more evaluation may be done to ascertain the medical history leading to reduced measurement errors in predictors.  For example, for a person diagnosed with lung cancer, there may be greater effort to clearly ascertain and document the patient's smoking history and environmental exposures, making it more likely to accurately record and retrieve this information.  Differential errors are more complex and difficult to account for in statistical analysis[1,2,3].  Berkson errors[1,4] are another type of error which arise when averages for subgroups of patients are assigned to individuals in each subgroup, e.g. average radiation exposure to workers in a facility where individuals have different actual exposures, or when missing values for patients are imputed by the estimated mean from a regression equation.

Many different models have been described for measurement error structures and to account for different error types.  Unfortunately, there does not seem to be a unifying way to analyze data affected by different types of measurement error.  Numerous computer programs are available for analyzing such data but the individual programs tend to be aimed at one type of error.  This absence of a general easy to use method and software is a further reason measurement error is often ignored at time of statistical analysis[1].

Most of the methods accounting for error essentially involve modelling a specific structure for the errors. These models can become complex and sensitive to deviations from the specified structure. Generally, to account for measurement error information must be available about the relation between the variables measured with and without error, or assumptions have to be made. Typically, information is gained from an internal subsample or external data. For the case with an internal validation subsample, Tong et al.[5], Wang and Wang[6] and Chen[7] and Chen and Chen[8], each considering a slightly different data structure, use a common method to derive estimates accounting for measurement error. Instead of trying to correlate variables measured with and without error directly, like for example in regression calibration, each of these papers, using data form the validation subsample, correlate regression coefficient estimates from a model involving variables subject to measurement error with regression coefficient estimates from a model involving variables not subject to measurement error. Based upon these correlations and the approximate multivariate normal distribution of the parameter estimates one can then obtain information on the regression coefficients for variables not subject to measurement error from the sample elements with only data for variables subject to measurement error. Without making assumptions about the error structures this approach is generally valid for different error types.

We describe in greater generality the approach used in these earlier works, unifying and extending the method to account for measurement error in the parametric regression setting, including linear, logistic and Cox regression, provided availability of an interval validation subsample.  As in earlier works the method does not require specification of the measurement error structure, and applies for non-differential, differential as well as Berkson errors.  Our new findings allow for measurement error in either the outcome or predictor variables or both, or all variables, and show for the Cox regression setting that the method accounts for measurement error in either or both the time to event and the indicator of event or censoring.  The method is not only generally applicable, but as newly implemented here using empirical methods, is simple to implement.  That is, one need not carry out extensive derivative and integration calculations, and need not program the expressions of the final integrals when working with our new results.  We present four example simulations showing how these estimates have at most a small bias and have markedly reduced variances compared to the estimates based only upon the variables without measurement error in the validation subsample, and one example with patient data.  We provide the open source R[9] package *meerva*, which calculates these estimates and requires only the data as input.

## METHODS

**Validation subsample**

Application of the proposed method is based upon a validation subsample with data for reference (true, ground-truth, gold standard) variables measured without error for outcome and predictor terms intended for use in the final model, where elements (patients) in the validation subsample are selected from the full sample by simple random sampling.  Further, the full sample includes surrogate variables, potentially measured with error, for the reference variables, outcome and predictors, intended for use in the final model. Some of the surrogate variables may be considered to be obtained without error and also included as reference variables in the validation subsample. These are considered as "perfect" surrogates. To avoid the degenerate case, at least one of the surrogates should (potentially) not be a perfect surrogate. The term surrogate is sometimes used to imply specific restrictions on measurement error[10]. Here we use the term to refer to any variable correlated with the reference variable[11,12].

**The Augmented estimator**

Assume we are working in the framework of a standard regression model, with sufficient sample size, so that the regression coefficient estimates are approximately multivariate normally distributed. This would include general linear models like linear, logistic and Poisson regression as well as the Cox proportional hazards regression models. Let $\hat{\beta}_{val}$ and $\hat{\gamma}_{val}$ be the regression coefficient estimates from the regression models based upon the reference and surrogate variables, respectively, from the validation subsample, and let $\hat{\gamma}_{ful}$ be the regression coefficient estimate from the model based upon the surrogate variables from the full sample (including the validation subsample). Then $\hat{\beta}_{val}$ will be approximately multivariate normally distributed with mean $\beta$, the true regression coefficient for the reference variables. Similarly, both $\hat{\gamma}_{val}$ and $\hat{\gamma}_{ful}$ are approximately multivariate normally distributed with mean $\gamma$, the true regression coefficient for the surrogate variables. It follows that the differences $(\hat{\beta}_{val} - \beta)$ and $(\hat{\gamma}_{val} - \hat{\gamma}_{ful})$ are approximately multivariate normally distributed with mean 0. Going one step further, the concatenated vector of $(\hat{\beta}_{val} - \beta)$ and $(\hat{\gamma}_{val} - \hat{\gamma}_{ful})$ will be approximately normal with mean 0, and variance-covariance $\begin{bmatrix} \Sigma & \Omega \\ \Omega^T & K \end{bmatrix}$ where $\Sigma$ and $K$ are the variance-covariance matrices for $(\hat{\beta}_{val} - \beta)$ and $(\hat{\gamma}_{val} - \hat{\gamma}_{ful})$ and $\Omega$ is the covariance matrix between $(\hat{\beta}_{val} - \beta)$ and $(\hat{\gamma}_{val} - \hat{\gamma}_{ful})$. Following Tong et al. and Wang and Wang, $(\hat{\gamma}_{val} - \hat{\gamma}_{ful})$ is observed, so conditional on it, by the (well known) properties of the multivariate normal distribution, $(\hat{\beta}_{val} - \beta)$ is approximately normal with mean $\Omega \, \mathrm{K}^{-1} (\hat{\gamma}_{val} - \hat{\gamma}_{ful})$ and variance $\Sigma - \Omega \, \mathrm{K}^{-1} \Omega^T$. This implies $\hat{\beta}_{val} - \Omega \, \mathrm{K}^{-1} (\hat{\gamma}_{val} - \hat{\gamma}_{ful})$ has mean $\beta$, and variance the same as $(\hat{\beta}_{val} - \beta)$, suggesting the basis of an estimator for $\beta$ using all available data. Let $\hat{\Sigma}$, $\hat{\Omega}$ and $\hat{\mathrm{K}}$ be estimates of variance-covariance terms $\Sigma$, $\Omega$ and $\mathrm{K}$. Replacing the variance-covariance matrices by their estimates yields the estimator for $\beta$

$$\hat{\beta}_{aug} = \hat{\beta}_{val} + \hat{\Omega} \, \hat{K}^{-1} (\hat{\gamma}_{ful} - \hat{\gamma}_{val}) \qquad (1)$$

and for the estimator variance

$$\widehat{var}(\hat{\beta}_{aug}) = \hat{\Sigma} - \hat{\Omega} \, \hat{K}^{-1} \hat{\Omega}^T \qquad (2)$$

Provided we can obtain estimates of the variance-covariance matrices, (1) and (2) together furnish a general tool for statistical inference, e.g. estimation, confidence intervals and tests, about the regression model for the reference variables using information from all data collected on both reference and surrogate variables.

Comparing (1) with the standard regression equation

$$\hat{Y} = \bar{Y} + \left(\frac{S_{XY}}{S_{XX}}\right)(X - \bar{X})$$

we see that (1) is of similar form as the standard linear regression equation but with $\hat{\beta}_{aug}, \hat{\beta}_{val}, \hat{\gamma}_{ful}, \hat{\gamma}_{val}$ in place of $\hat{Y}, \bar{Y}, X$ and $\bar{X}$. Conceptually then $\hat{\beta}_{aug}$ can be thought of as the solution regressing $\hat{\beta}_{val}$ on $\hat{\gamma}_{val}$ and making an estimate based upon the observed $\hat{\gamma}_{ful}$. Here, $\widehat{\Omega}$ is the correlation between $(\hat{\gamma}_{ful} - \hat{\gamma}_{val})$ and $\hat{\beta}_{val}$ instead of between X and Y (as in S$_{XY}$), and $\widehat{K}$ is the variance of $(\hat{\gamma}_{ful} - \hat{\gamma}_{val})$ instead of X (as in S$_{XX}$). Note, in the above derivation we can use $\hat{\gamma}_{non}$ in place of $\hat{\gamma}_{ful}$, where $\hat{\gamma}_{non}$ is the regression parameter estimate based upon the surrogate variable data not in the validation subsample. Then the formulation parallels that of linear regression even more closely with $\hat{\gamma}_{non}$ being independent of $\hat{\gamma}_{val}$, and the correlation matrix in (1) involving only data from the validation subsample. Since the formula for these estimates may be interpreted as an estimate based upon the validation subsample augmented by information from the full sample, Wong et al. describe estimators of this form as "augmented" estimators. We continue with this naming convention.

**Estimating variance-covariances**

For a regression model based upon the surrogate variables, the usual model assumptions may not hold even if they hold for the reference variables[3]. For example, when the logistic model assumptions hold for the reference variables, if the surrogate outcome variable is ascertained with misclassification error, then the probability for the outcome variable is bounded away from 0 and 1 and the logistic model no longer holds. Model assumptions in general may be violated for the models fit using surrogate variables. Therefore, robust methods for estimating variances should be used. We consider numerical methods like the jackknife, sandwich estimators like the infinitesimal jackknife and the bootstrap to obtain robust variance covariance estimates.

Tong et al.[5], Wang and Wang[6], Chen[7] and Chen and Chen[8] all use versions of the infinitesimal jackknife. Variance estimates based upon the infinitesimal jackknife have been previously generally described[13-Efrom, 16-Freedman] for many parametric (regression) models.  The infinitesimal jackknife works by approximating the change in the parameter estimate when leaving out one observation from the analysis dataset. These changes are termed DFbeta and are recalculated, once for each observation. Further, the sums of the cross products of these DFBeta's robustly estimates the variances of the original parameter (here regression) estimate.   In particular, if DFBeta(i) is the change in the regression coefficient estimate when leaving out observation i, and expressed as a column vector, then

$$\sum_i \text{DFBeta}(i)\text{DFBeta}(i)^T - (1/n)\left(\sum_i \text{DFBeta}(i)\right)\left(\sum_i \text{DFBeta}(i)\right)^T$$

can be used estimate the individual parameter variances and covariances $\Sigma$, $\Omega$ and $K$. The complicating factor for augmented estimates is that the term $(\hat{\gamma}_{val} - \hat{\gamma}_{ful})$ involves a difference between two estimators and thus the difference between two DFBeta's, and for some indices i, $\hat{\gamma}_{val}$ does not change, making calculations slightly more tedious. In this regard the augmented estimate based upon the difference $(\hat{\gamma}_{val} - \hat{\gamma}_{non})$ is slightly easier to work with because $\hat{\gamma}_{val}$ and $\hat{\gamma}_{non}$ are derived from nonoverlapping subsamples and thus independent.  The numerics of infinitesimal jackknife are well established in practice and DFBeta's are frequently, if not typically, provided by statistical softwares for regression analysis.  Therefore, derivation of DFBetas for variance estimation as described in the earlier works on augmented estimation are often already available and need not be rederived.  Interestingly, though, there exist different approximations for the DFBeta's and at least for the logistic regression framework the approximation described by Tong at al. outperforms some of the commonly used approximations.

The jackknife works by calculating the parameter estimate when leaving out one observation from the analysis dataset, in contrast with the infinitesimal jackknife which works with an approximation of the change. In general, the jackknife estimator of variance will be slightly more accurate than the infinitesimal jackknife estimator but is also more computationally intensive. For $\hat{\theta}_{(i)}$ the estimate calculated leaving out observation i, the jackknife estimator is

$$\sum_i \hat{\theta}_{(i)} \hat{\theta}_{(i)}^T - (1/n)\left(\sum_i \hat{\theta}_{(i)}\right)\left(\sum_i \hat{\theta}_{(i)}\right)^T$$

With $\hat{\theta}_{(i)}$ serving as a generic term for $\hat{\beta}_{val}$, $(\hat{\gamma}_{val} - \hat{\gamma}_{ful})$, $(\hat{\gamma}_{val} - \hat{\gamma}_{non})$ or the concatenation of $\hat{\beta}_{val}$ and $(\hat{\gamma}_{val} - \hat{\gamma}_{ful})$, or $\hat{\beta}_{val}$ and $(\hat{\gamma}_{val} - \hat{\gamma}_{non})$, the jackknife can be used to estimate the variances and covariances Σ, Ω and K needed for the calculation of $\hat{\beta}_{aug}$ and $\widehat{var}(\hat{\beta}_{aug})$. Note, for the jackknife we did not take a difference between the leave-out parameter estimate and the original estimate like with the DFBeta's but calculated cross products using the leave out estimates directly. Because we subtract out the (scaled) product of the averages we get the same results either way. Despite their similarity the conventions of describing the infinitesimal jackknife and the jackknife are sometimes different.

    The bootstrap is another method that can be used to provide potentially even more robust variance and covariance estimates but is computationally even more intensive than the jackknife. In our simulation studies we have found the infinitesimal jackknife and jackknife to perform very well and have not had the need to use bootstrap variance estimates. We refer the reader to Efron[13] for a discussion on the bootstrap.

**SIMULATIONS**

We simulated multiple scenarios for yes/no (1/0) outcomes from the logistic model, time to event data from the Cox proportional hazards regression model, and quantitative outcomes from the linear model, with duplicate surrogates and with surrogates without random error. Here, we present simulations demonstrating our general findings. Each of these simulations includes 4 predictors in the regression equation plus an intercept term for the logistic and linear regression models, and each is based upon a full sample size of 4000 and validation subsample of size 400. 1000 datasets were simulated for each scenario. In the first 3 examples, we include 2 yes/no (0/1) class predictors and 2 numerical predictors, and for the fourth example 4 numerical predictors.

**Example 1 – Differential misclassification of outcome in logistic regression model**

We first simulated logistic regression model data as described by Tong where the yes/no outcome is subject to differential misclassification. From **Figure 1** the augmented estimates were less variable than the estimates based upon the reference variables in the validation subsample alone, and were less biased than estimates based upon the surrogate variables from the full dataset, resulting in the augmented estimates having the smallest root mean square errors (**Table 1**). Observed coverage probabilities for the augmented estimate's 95% confidence intervals ranged from 94.5% to 95.7% and were all indistinguishable from 0.95. Coverage probabilities for the full sample surrogate based ($\hat{\gamma}_{ful}$) confidence intervals ranged from 0.0% to 3.2%. Note, Tong et all show how the augmented estimates may perform somewhat worse than a likelihood based approach when error model assumptions[15] hold, or much better in case of measurement error misspecification.

**Example 2 – For time to event data, measurement error in time to event, differential misclassification of event (vs. censoring), differential misclassification of a predictor, and surrogate non-linear in the reference predictor and with non-constant error variance.**

The augmented estimates based upon a Cox regression model were on average very close to the actual true regression coefficients used to generate the data and had reduced variability compared to the estimates based upon the reference variables from the validation subsample (**Figure 2**). Estimates based upon the surrogate variables in the full dataset had much smaller variability but were rather biased, leading to significantly greater root mean square errors (**Table 1**). None of the observed coverage probabilities for the augmented estimate's 95% confidence intervals were significantly lower than 0.95, but the coverage probabilities for the full sample surrogate based ($\hat{\gamma}_{ful}$) confidence intervals ranged from 0.8% to 93.1%. Note, for the Cox model, the intercept is absorbed into the baseline survival function, and no β for the intercept is derived or displayed in Figure 2.

**Example 3 – Differential measurement errors in quantitative outcome, differential misclassification of a predictor, and one numerical reference with <u>two surrogates</u>, each with measurement error.**

We have encountered cases like this with multiple surrogates for the same reference when multiple laboratory tests aim to measure the same as a more invasive or costly reference method. The augmented estimates based upon a linear model had minimal bias, and were less variable than estimates based on the reference variables alone in the validation subsample (**Figure 3**).  Regression coefficient estimates based upon the surrogate variables mostly had reduced standard deviations but had larger bias and root mean square error, especially for the dual surrogates ($\beta_4$ and $\beta_{4d}$ in **Figure 3** and **Table 1**). Coverage probabilities for the augmented estimate's 95% confidence intervals were indistinguishable from 95%.

**Example 4 – The linear model setting with all surrogates as linear transformations of scale with no other error**

We next simulated datasets according to a linear model with normal errors, where the surrogate variables were simple linear transformations in scale of the reference variables.  Simplistically this could happen if for example surrogates were recorded in centimeter instead of inch, kilogram instead of pound or centigrade instead of Fahrenheit. The augmented estimates based upon a linear model were on average very close to the actual true regression coefficients used to generate the data with standard deviations and root mean square errors about those of the estimates based upon the reference variables in the validation subsample divided by the square root of 10 (**Figure 4** and **Table 1**) .  This is expected with the full sample having size 10 times greater than that of the validation subsample, and estimate variances being inversely proportional to sample size.  This shows how the augmented method can efficiently capture information on the reference variables contained in the surrogate variables. The observed coverage probabilities for the augmented estimate's 95% confidence intervals were all indistinguishable from 0.95.

**EXAMPLE ANALYZING ELECTRONIC HEALTH RECORD DATA**

For a current study (approved by the Mayo Clinic Institutional Review Board) we are describing risk of joint infection following implant of a prosthetic joint, knee or hip. We have surrogate variables for comorbidities based upon natural language processing (NLP) as well as coded variables, both of which may be in error, and randomly selected roughly 7% of the patients in the original dataset for manual chart review of the medical records to determine reference variables for these comorbidities. For this example, we consider a reduced set of predictor variables and describe the augmented estimates for the reference variables, regression model built for the reference variables based upon the validation subsample alone and a regression model for the surrogate variables based up upon the whole data set are given in **Table 2**. In this table chf denotes chronic heart failure, de dementia, ref reference, nlp.sur an NLP surrogate and icd.sur and ICD (International Classification of Diseases) code surrogates. This example using actual clinical data shows how the standard errors (SE) are drastically reduced for the augmented estimates compared to the estimates based upon the validation sample alone, and much closer to the SEs one would expect had there been an increase in sample size for the reference variables to that of the whole dataset. Without imposing any type of measurement error structure in the model the augmented estimates will be more robust than the estimates based upon the surrogate variables. Further, for this example there are two surrogates for each reference comorbidity variable, making it difficult know how one might even compare reference and surrogate variable coefficient estimates. Here, ICD provides a much better surrogate for chronic heart disease and NLP provides a much better surrogate for dementia. (For the final study analysis we will consider additional variables and more of the medical complexities of PJI.)

**DISCUSSION**

We describe a general method accounting for measurement error, as can occur in epidemiological data, for the case when surrogate variables exist in the full sample and reference variables are available for analysis from a randomly selected subsample of the full sample. The method is general, simple, computationally modest and statistically robust, accounting for differential, non-differential and Berkson measurement errors, yielding augmented estimates better than estimates based upon either the surrogate variables from the whole sample or the reference variables from the internal validation subsample. The proposed method also accounts for the case where individual reference predictors may have multiple surrogates. The method can be applied quite generally to the parametric regression setting including linear, logistic and Cox regression. We extended earlier works[5,6,7,8] by showing how the augmented estimates can be used when there are measurement errors in both outcome and predictors, and for the Cox regression setting where the measurement errors may involve both the time to event and the event indicator variables. We also extend earlier works by showing that augmented estimates can be used for numerous different regression model frameworks by either using already existing numerical routines for DFBeta calculation, or using numerical methods like the jackknife to derive estimates without the user having to carry out pages of derivative and integration calculations. Simulations showed how the augmented estimates have statistical properties as designed in terms of bias, confidence coverage probabilities and decreased mean square error compared to estimates based upon either the surrogate variables or reference variables alone.

In earlier works the augmented estimator was expressed as

$$\hat{\beta}_{aug} = \hat{\beta}_{val} - \hat{\Omega}\, \hat{K}^{-1} (\hat{\gamma}_{val} - \hat{\gamma}_{ful})$$

While this is correct, we think it masks some inherent aspects of the estimator. When we reverse the order of the two estimators for the surrogate data then the estimator takes on a form like that in linear regression as described in **Methods**. It is easier to use, check and be confident in a model when one better understands how it works.

The method described here can be generalized in yet more ways. As described in **Methods**, the validation subsample is drawn as a simple random sample. Elements of the subsample though can be drawn with unequal probabilities in which case one can weight by inverse sampling probabilities to obtain appropriate estimates of the correlations needed for the calculation of the augmented estimates. These weights too can be used for propensity score adjustment. Tong et al.[5] describe this as an area for future research. Wang and Wang[6] acknowledge that inverse weights may be incorporated following earlier works. We observe that wights may be incorporated into the matrix formulae for the surrogate estimate formulae in the usual manner and have incorporated weights in this manner as an option in our accompanying software. If there are multiple data points for some or all elements (persons) in a dataset then one can account for these clusters of dependent observations in the derivation of the robust variance-covariance estimates. (We did this for the example involving EHR data and this too is incorporated into our accompanying software.) The surrogate variables may be not only measured with error but of a different structure than the reference variables. For example, the reference outcome could be a yes/no outcome and analyzed using a logistic model, while the surrogate outcome could be determined as the estimated classification probability from the penultimate step of a natural language processing recursive neural network based upon the medical record, or logit of this estimated probability.  In this case one may still estimate the variance-covariances between the regression parameter estimates from the models for the reference and surrogate variables to derive augmented estimates for the reference variable association.  We have not seen this described before.  As noted by Chen and Chen[8] there may be fewer surrogates than reference variables, and the augmented estimates may be used to address this case of missing variables. More extensions can be imagined.

  In general, robust variances and covariances for parameter estimates can be obtained using resampling methods.  By their basic nature, these resampling variance estimates depend on the independence of the sample elements, and as applied here a validation subsample selected by simple random sampling.  However, the calculation of the augmented estimates does not depend on exact correctness of the specified model, for example linearity of the response curve or constancy of variances[13-Efron].  Numerical methods such as the jackknife though, like the underlying regression models, are sensitive to extreme values and so care should be taken in case of heavy tailed distributions and transformations should be considered[14-Miller].  Whereas the interpretability of the final model will depend on the appropriateness of the model as applied to the reference variables[16-Freedman], the stability of the estimates and the coverage probabilities of the confidence intervals do not depend on the usual model-based assumptions holding for the surrogate-based models.

One limitation of the method is it cannot recover information that is not there. From the formulae (1) and (2), the observation by Tong[5] et al. for their setting is equally applicable here: when surrogates are uncorrelated with the reference variables, then the augmented estimates will do no better than estimates based upon the reference variables in the validation subsample alone.  Similarly, when the surrogates are perfectly correlated with the reference variables, then the variance-covariance for the augmented estimated will be about that expected from the increase in sample size from the validation to the full sample.  In general, the amount of adjustment in the augmented estimate and decrease in variance depends on the strength of the correlation between the reference and surrogate variables, and the imposed (squared multiple) correlation between $\hat{\beta}_{val}$ and the $(\hat{\gamma}_{val} - \hat{\gamma}_{ful})$ difference[8-Chen and Chen].

As shown by Tong for their setting where only the outcome is obtained with misclassification error and analyzed using logistic regression, the augmented estimates may not do as well as a likelihood approach when the error structure is known but may do much better when error model assumptions used in the likelihood approach do not hold.  Also, when the validation sample size is limited, it may be difficult to ascertain or even evaluate a specific error structure for measurement or misclassification errors[4-Brakenhoff].  The augmented estimates based upon surrogate variables in the full sample and reference variables from a random subsample, is robust and generally applicable.

For the linear model case Tao et al[17] recently described a method accounting for measurement error in both outcomes and predictors.  Their model-based approach though assumes a type of non-differential error structure by requiring that the errors for both outcome and predictors be independent of the errors in the underlying regression model and so does not generally account for differential measurement or misclassification errors.  This is similar to the assumptions of the paper by Yi et al[18] who consider a more general model structure, o errors in the outcome, but again independences of measurement error form the underlying model error.  (The Yi paper is the basis for the augSIMEX R-package.) Like for the logistic regression setting, model-based estimates may outperform augmented estimates, but if assumptions for the model based method do not hold then the augmented estimates can outperform the model-based estimates.  Other limitations of the Tao method are it only considers a continuous outcome, and does not account for multiplicative errors in the outcome variable.  The approaches by both Tao and Yi do not seem to apply for cases where the model structure may be different between the models based upon reference and surrogate variables.  The method we describe accounts for all of these factors and thus has broader applicability.  A complication of the method by Tao is that when there are multiple numerical surrogates the number of spline terms used to describe the individual surrogates, and their potential interaction or product terms, may lead to a large increase in number of parameters far exceeding the number of parameters in the augmented method, reducing the efficiency of the model-based estimates compared to the augmented estimates.  However, if the measurement error for multiple predictors only involves misclassification of yes/no variables then dimensionality of Tao's model may remain modest and the estimates may continue to perform well.  When they can be used and their assumptions hold the model-based methods should outperform the augmented estimates, but this benefit may no longer hold when there are multiple quantitative surrogates.

This method to account for measurement error based upon an internal validation subsample, without modeling any specific measurement error, is general, simple, computationally modest, statistically robust, and accounts for differential, non-differential and Berkson measurement errors, yielding estimates better than estimates based upon either the surrogate variables from the whole sample or the reference variables from the internal validation subsample.  From simulation studies we saw the improvements in estimation predicted by the model and for confidence interval coverage probabilities to be very close to the nominal levels demonstrating the applicability of the method.  The method can also account for inverse probability of sampling weighting and multiple, dependent observations from individual patients, addressing more involved study designs.  This method presents a valuable option for analysis of data with measurement error when an internal validation subsample may be obtained.  We provide the open source R[9] package *meerva*, which calculates these estimates and requires only the data as input, making this method practical to implement.

**Table 1. Root Mean Square Errors for Estimates from Simulation Studies.**

The (square) root mean square errors of estimates are calculated from the simulated data displayed in Figures 1-4. Values are smaller for the augmented estimates ($\beta_{aug}$) than for those derived from the reference variables in the validation subsample ($\beta_{val}$), the degree to which is dependent on the strength of the correlation between the reference and surrogate variables. Values for the augmented estimates are generally smaller than those for the estimates based upon the surrogate variables in the full sample ($\gamma_{ful}$) due to bias, but in some cases they are larger. (A broken clock is exactly correct twice a day.)

|  | $\beta_0$ (Int.) | $\beta_1$ | $\beta_2$ | $\beta_3$ | $\beta_4$ |
|---|---|---|---|---|---|
| Logistic Regression — Example 1 | | | | | |
| $\beta_{aug}$ | 0.2508 | 0.2150 | 0.2755 | 0.1960 | 0.1066 |
| $\beta_{val}$ | 0.3246 | 0.2760 | 0.3524 | 0.2161 | 0.1317 |
| $\gamma_{ful}$ | 0.3465 | 0.3668 | 0.4305 | 0.8050 | 0.1459 |
| Cox Regression — Example 2 | | | | | |
| $\beta_{aug}$ | . | 0.0901 | 0.0843 | 0.0568 | 0.0367 |
| $\beta_{val}$ | . | 0.1294 | 0.1582 | 0.0817 | 0.0679 |
| $\gamma_{ful}$ | . | 0.1964 | 0.0549 | 0.1182 | 0.0503 |
| Linear Regression — Example 3 | | | | | |
| $\beta_{aug}$ | 0.0461 | 0.0648 | 0.0425 | 0.0167 | 0.0171 |
| $\beta_{val}$ | 0.1198 | 0.1011 | 0.1251 | 0.0506 | 0.0511 |
| $\gamma_{ful}$ | 0.0709 | 0.1539 | 0.0395 | 0.0156 | 0.2076 |
|  |  |  |  |  | 0.2535 |
| Linear Regression — Example 4 | | | | | |
| $\beta_{aug}$ | 0.0157 | 0.0160 | 0.0155 | 0.0159 | 0.0156 |
| $\beta_{val}$ | 0.0501 | 0.0488 | 0.0525 | 0.0518 | 0.0498 |
| $\gamma_{ful}$ | 0.0616 | 0.0261 | 0.0447 | 0.0895 | 0.0588 |

**Table 2. Estimates from a dataset involving 58,574 records, 28,069 patients and 28069 records and 1882 patients in a validation subsample, based upon the augmented method and an underlying logistic regression model.**

```
Estimates for Beta using beta_aug
     (reference variables augmented with surrogates)
              estimate    se     lcl      ucl      z        p
(Intercept)   -2.989    0.234  -3.448   -2.529  -12.749  3.15e-37
chf.ref        0.633    0.161   0.317    0.949    3.925  8.66e-05
de.ref         0.586    0.167   0.260    0.913    3.520  4.32e-04
age_10        -0.873    0.329  -1.152   -0.229   -2.658  7.87e-03
male           0.173    0.077   0.021    0.325    2.234  2.59e-02
hip           -0.097    0.078  -0.249    0.055   -1.252  2.10e-01
surg.primary  -1.472    0.078  -1.625   -1.319  -18.858  2.51e-79

 Estimates for Beta using beta_val (reference variables alone)
            estimate    se      lcl      ucl      z        p
(Intercept)  -3.849    0.882   -5.578   -2.120  -4.364   1.28e-05
chf.ref       0.749    0.381    0.003    1.496   1.967   4.92e-02
de.ref       -0.958    0.990   -2.899    0.983  -0.967   3.34e-01
age_10        0.116    1.181   -2.199    2.430   0.098   9.22e-01
male          0.312    0.308   -0.292    0.916   1.012   3.11e-01
hip          -0.280    0.320   -0.906    0.347  -0.875   3.81e-01
surg.primary -1.570    0.312   -2.182   -0.958  -5.030   4.92e-07

 Estimates for Gamma using gamma_ful (surrogate variables alone)
            estimate    se      lcl      ucl      z        p
(Intercept)  -2.931    0.230   -3.381   -2.480  -12.761  2.72e-37
chf.nlp.sur   0.074    0.134   -0.189    0.337    0.551  5.81e-01
chf.icd.sur   0.814    0.146    0.528    1.099    5.591  2.25e-08
de.nlp.sur    0.504    0.161    0.189    0.820    3.127  1.76e-03
de.icd.sur   -0.166    0.221   -0.599    0.266   -0.754  4.51e-01
age          -0.925    0.319   -1.551   -0.300   -2.899  3.74e-03
male          0.158    0.075    0.010    0.306    2.087  3.69e-02
hip          -0.089    0.076   -0.238    0.059   -1.182  2.37e-01
surg.primary -1.455    0.076   -1.603   -1.306  -19.206  3.29e-82
```

**Figure 1. Augmented estimates compared to estimates based upon the reference variables or surrogate variables alone for a logistic regression model with measurement error**

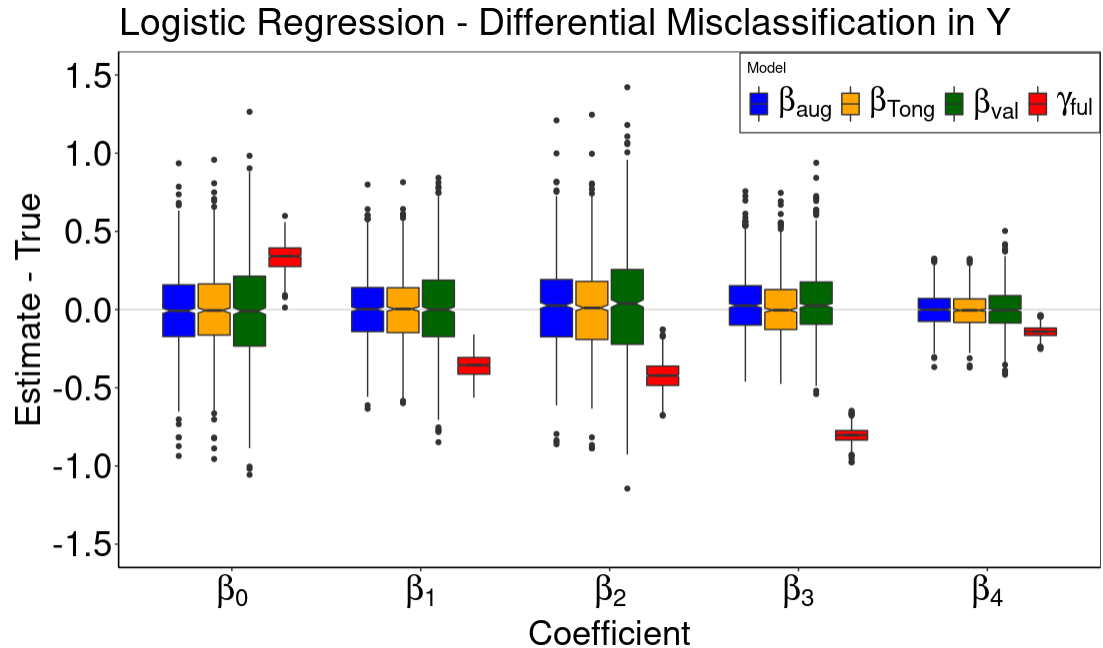

Augmented estimates were calculated based upon each of 1000 randomly generated datasets from a logistic regression model where $P(Y=1) = (1+\exp(-X\beta))^{-1}$ with $\beta = (\beta_0, \beta_1, \beta_2, \beta_3, \beta_4) = (-0.5, 0.25, 1.00, 1.90, 0.33)$. $\beta_0$ and $\beta_1$, $\beta_2$, $\beta_3$ and $\beta_4$ are the regression coefficients for the intercept and slope terms in the regression model. $X_1$ and $X_2$ are generated as Yes/No random variables with probability 0.4 for $X_1$ being 1 (Yes), and probability 0.8 for $X_2$. $X_3$ and $X_4$ are generated as normal random variables with mean 0 and variance 1. The value for $\beta$ is taken from Tong et al.[5], where Y represents the outcome, $X_1$ and $X_2$ represent exposure and race, and $X_3$ and $X_4$ represent age and BMI, but $X_3$ and $X_4$ were rescaled to have a standard deviation of 1. The surrogate for Y (outcome) is differentially misclassified with sensitivity and specificity of 0.85 and 0.90 when $X_1 = 1$, and of 0.90 and 0.85 when $X_1 = 0$. Each dataset has a full sample size of 4000 and validation subsample size of 400.

This and the following figures show estimates for the regression coefficient estimates obtained using the augmented method ($\beta_{aug}$) minus the true regression coefficient used to generate the data. For comparison we also display estimates (minus the true regression coefficients) derived from the validation subsample reference variables ($\beta_{val}$) and the full sample surrogate variables ($\gamma_{ful}$). In the boxplots, boxes show the interquartile range, whiskers extend up to 1.5 interquartile ranges from the boxes, individual points are displayed beyond that, and notches show approximate 95% confidence intervals for the median.

**Figure 2. Augmented estimates for a Cox regression model with measurement error**

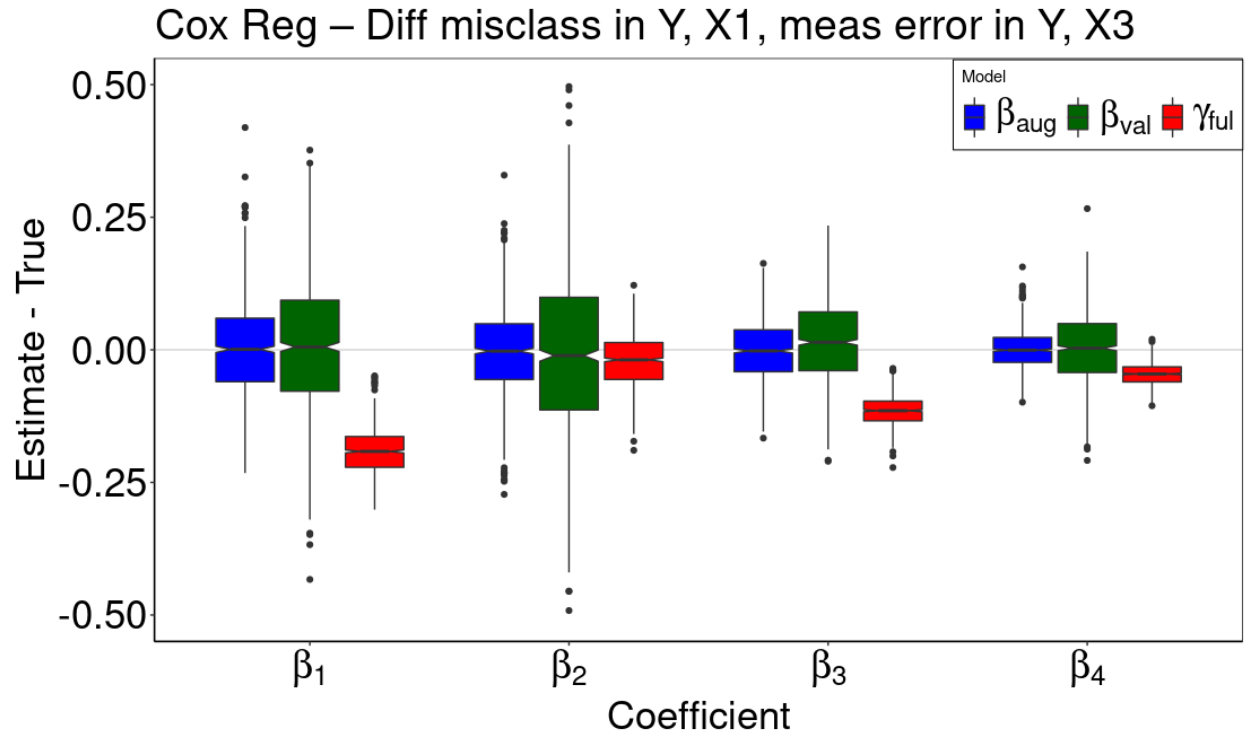

Augmented estimates were calculated based upon each of 1000 randomly generated datasets from a Cox regression model where exponential failure times, T, were generated with rate Xβ and β = (-0.5, 0.5, 0.2, 1, 0.5). Exponential censoring times, C, were generated with rate 0.3, and failure times were censored at the minimum of C and 3. Observed are Y = min(T, C, 3) and E (event outcome), where E = 1 if T<= min(C, 3), else E=0. $\beta_0$, $\beta_1$, $\beta_2$, $\beta_3$ and $\beta_4$ are the regression coefficients in the Cox model. $X_1$, $X_2$, $X_3$ and $X_4$ are generated as for Figure 1. The surrogate for T was T times the exponential of normal random variable with standard error 0.05. The surrogate for E (event outcome) is differentially misclassified with sensitivity and specificity of 0.90 and 0.95 when $X_1$ = 1 and of 0.95 and 0.90 when $X_1$ = 0. The surrogate for $X_1$ is differentially misclassified with sensitivity and specificity of 0.95 and 0.95 when E = 1 and of 0.90 and 0.90 when E = 0. The surrogate for $X_3$ is derived by first taking max(0, $(X_3+2))^2$, scaling these values to have a standard deviation of 0.9, and then adding a normal error with standard deviation max(0.05, $0.05(X_3+2)$). This surrogate is roughly a quadratic in $X_3$ and has variance increasing in $X_3$.

**Figure 3**. **Augmented estimates for a linear regression model with measurement error including duplicate surrogates**

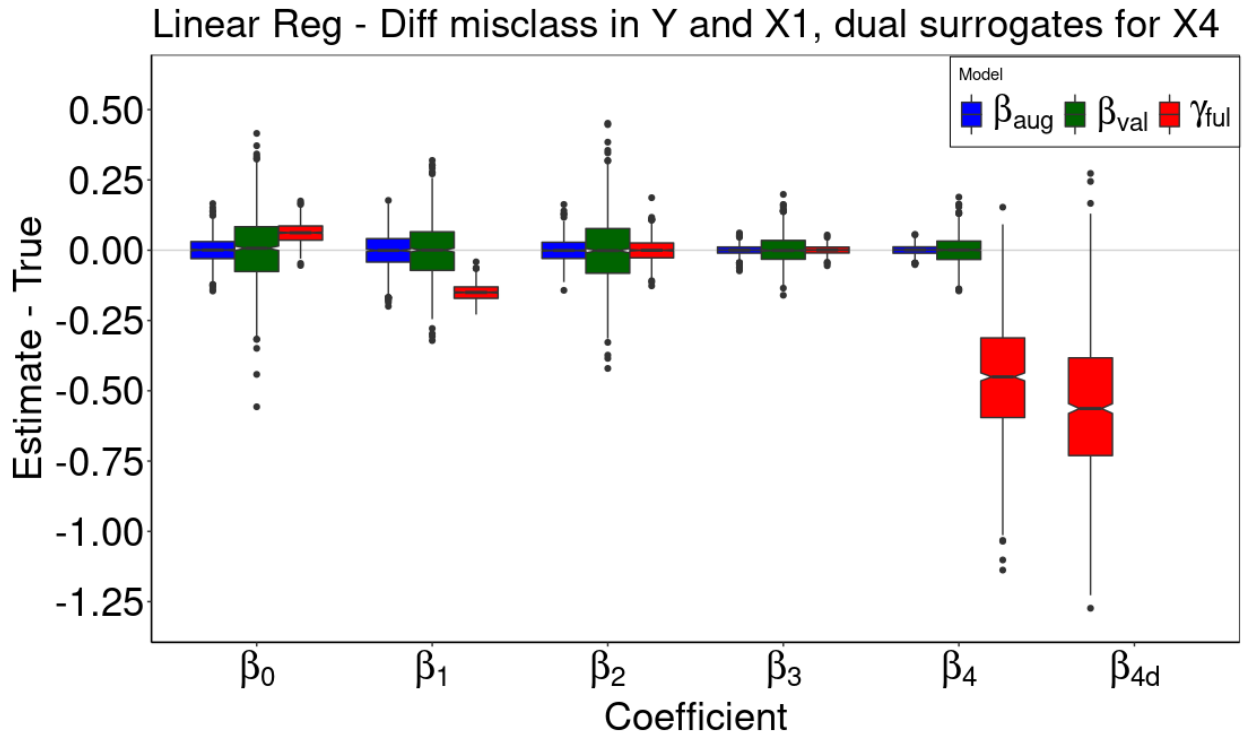

Augmented estimates were calculated based upon each of 1000 randomly generated datasets from a linear regression model where mean for the outcome Y is Xβ, β = (-0.5, 0.5, 0.2, 0.5, 1) and model errors are independent normals with variance 1. $β_0$, $β_1$, $β_2$, $β_3$ and $β_4$ are the regression coefficients for the intercept and for $X_1$, $X_2$, $X_3$ and $X_4$. The surrogate for Y has differential error of -0.5 + e when $X_1$ = 1 and 0.5 + e when X1 = 0 where e is a normal random variable with standard deviation 0.1. $X_1$, $X_2$, $X_3$ and $X_4$ are generated as for Figure 1. The surrogate for $X_1$ is differentially misclassified with sensitivity and specificity of 0.95 and 0.91 when the model error is positive, and of 0.90 and 0.90 when the model error is negative. There are two surrogates for $X_4$, the first being 1.1 $X_4$ + $e_1$, and the second 0.9 $X_4$ + $e_2$, where $e_1$ and $e_2$ are independent normal with standard deviation 0.05. We let $β_{4d}$ denote the regression coefficients for the duplicate surrogate.

**Figure 4**. **Augmented estimates for a linear regression model with surrogates without random error**

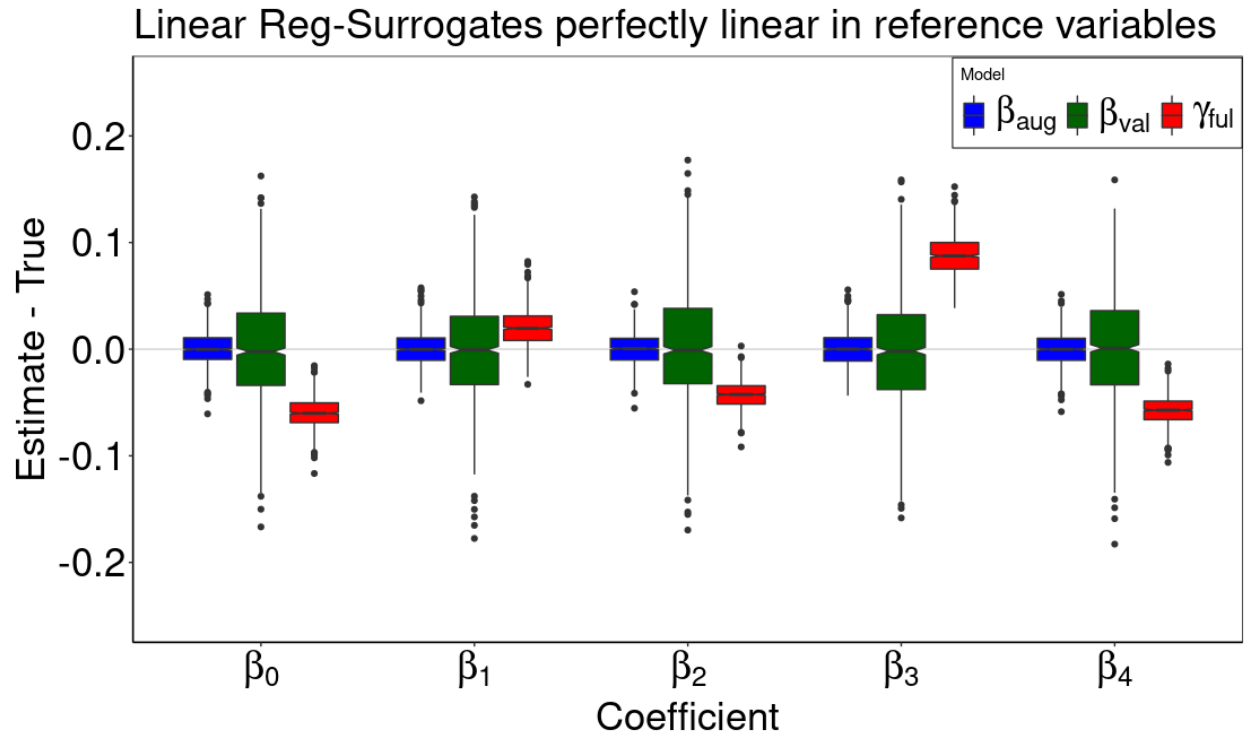

Augmented estimates were calculated based upon each of 1000 randomly generated datasets from a linear regression model where mean for the outcome Y is X$\beta$, $\beta$ = (2, 0.2, 0.3, 0.7, 0.4) and model errors are independent normals with variance 1. $\beta_0$ and $\beta_1$, $\beta_2$, $\beta_3$, $\beta_4$ are the regression coefficients for the intercept and for $X_1$, $X_2$, $X_3$ and $X_4$. $X_1$, $X_2$, $X_3$ and $X_4$ are generated as normal random variables with mean 0 and variance 1. All predictors are quantitative and their surrogates are linear in the reference variables ($X_1$, $X_2$, $X_3$, $X_4$) with no random error component: $Y_s$ = 0.14 + 0.9 Y, $X_{1s}$ = 0.82 $X_1$, $X_{2s}$ = 1.1 $X_2$, $X_{3s}$ = 0.80 $X_3$ and $X_{4s}$ = 1.05 $X_4$.

**Figure 1**

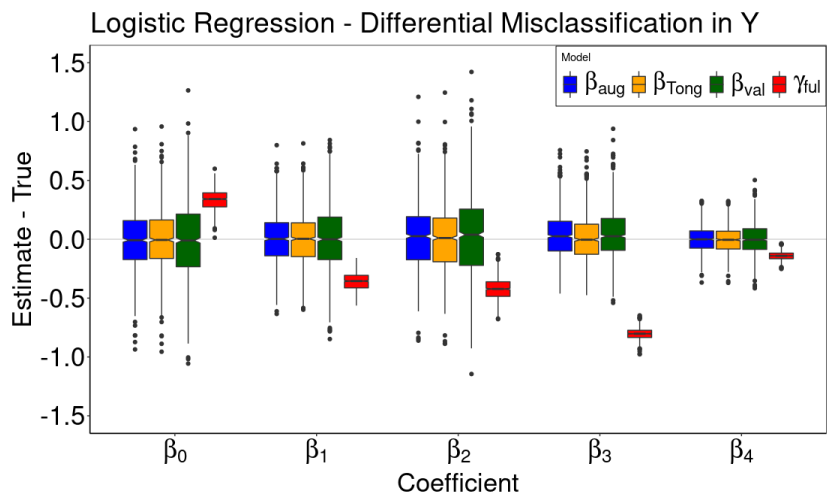

**Figure 2**

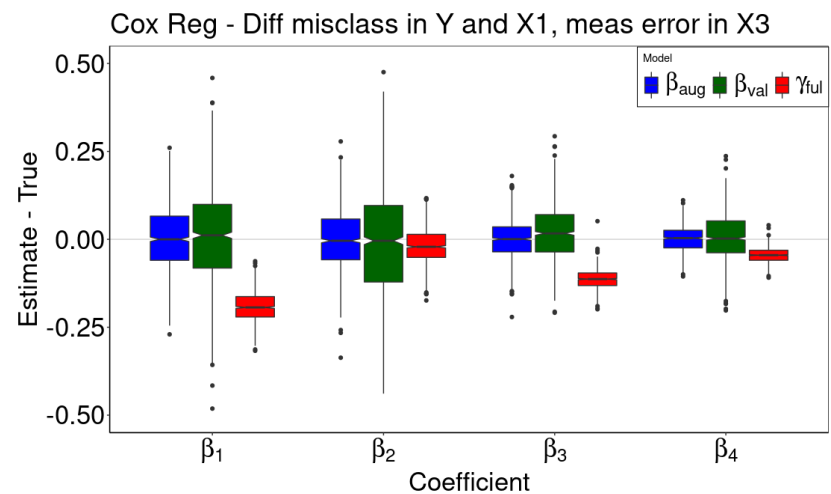

**Figure 3**

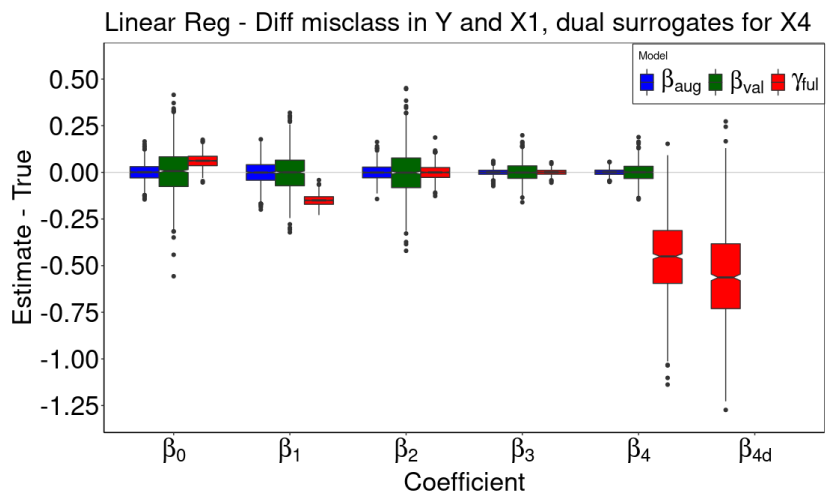

**Figure 4**

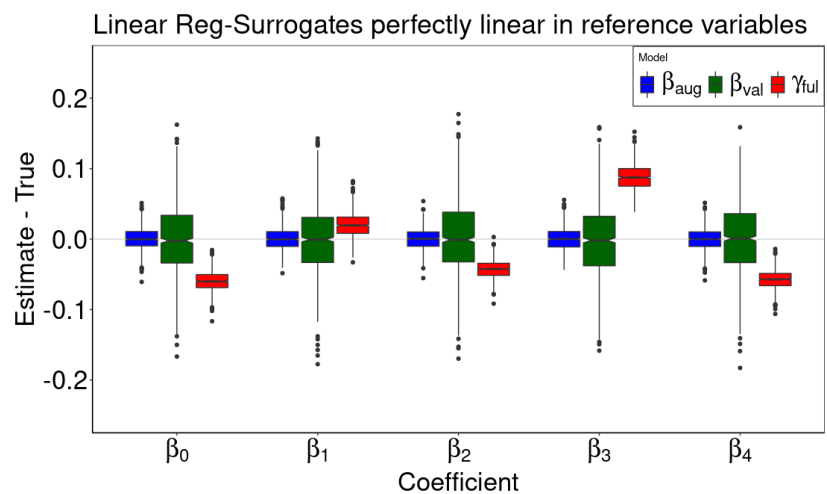